\documentclass{elsart3}
\usepackage{times}
\usepackage{amssymb}
\usepackage{amsfonts}
\usepackage{eucal}
\usepackage{revsymb}
\usepackage{graphicx}
\usepackage{epsfig}

\begin{document}

\begin{frontmatter}
\title{ Singular Coexistence-curve Diameters: Experiments and Simulations}

\author{Young C.\ Kim and Michael E.\ Fisher\corauthref{cor1}}
\corauth[cor1]{Corresponding Author: xpectnil@ipst.umd.edu}
\address{Institute for Physical Science and Technology, University of Maryland, College Park, Maryland 20742 USA}

\begin{abstract}
Precise calculations of the coexistence-curve diameters of a hard-core square-well (HCSW) fluid and the restricted primitive model (RPM) electrolyte exhibit marked deviations from rectilinear behavior. The HCSW diameter displays a $|t|^{1-\alpha}$ singularity that sets in sharply for $|t|\equiv |T-T_c|/T_c\lesssim 10^{-3}$; this compares favorably with extensive data for $\mbox{SF}_6$, also reflected in C$_2$H$_6$, N$_2$, etc. By contrast, the curvature of the RPM diameter varies slowly over a wide range $|t|\lesssim 0.1$; this behavior mirrors observations for liquid alkali metals, specifically Rb and Cs. Amplitudes for the leading singular terms can be estimated numerically but their values cannot be taken literally.
\end{abstract}

\end{frontmatter}

\section{Introduction}
\label{sec1}
Coexistence curves of gas-liquid phase separation in single-component fluids have been investigated extensively for more than a century via both experiment and theory. In particular, the van der Waals (vdW) equation of state has been successful in describing the experimental data at a semiquantitative mean-field level \cite{ani:sen}; however, the shape of the coexistence curve predicted by the vdW equation of state differs significantly from observation. Nevertheless, one aspect of the vdW equation of state is that the diameter of the coexistence curve, $\rho_{\mbox{\scriptsize diam}}(T)$, namely the mid-points of the coexistence-curve boundary in the density-temperature plane, exhibits a linear behavior with temperature, $T$, as the critical point, $T_c$, is approached; this is the Law of the Rectilinear Diameter, and is typically observed experimentally to be valid with a fairly high degree of precision. 

Several models for fluids, however, have been advanced that violate the law \cite{wid:row,hem:ste,mer1}. These models suggest that the diameters of real fluids should exhibit an entropy-like singularity varying as $|t|^{1-\alpha}$ near criticality where $t\equiv (T-T_c)/T_c$ and $\alpha\simeq 0.109$ is the critical exponent for the specific heat; the slope of the diameter was thus expected to diverge at criticality in the same fashion as the specific heat. The traditionally accepted {\em scaling theory} of the critical region, which incorporates mixing between the temperature $T$ and the chemical potential $\mu$ in the scaling fields (but does {\em not} include the pressure), also generates this entropy-like singularity \cite{gre:coo:sen,mer:reh}. However, the recent discovery of a Yang-Yang anomaly \cite{fis:ork,ork:fis:ust} in fluid criticality turns out to entail a {\em further} singular term in the diameter proportional to $|t|^{2\beta}$, where $\beta\simeq 0.326$ is the critical exponent for the order parameter \cite{fis:ork,kim:fis:ork}. And such a term also arises in other classes of exactly soluble model \cite{fis:ork,fel:fis}. In fluid criticality, one normally has $2\beta\simeq 0.65 < 1$$-\,$$\alpha$ so that the $|t|^{2\beta}$ term dominates the $|t|^{1-\alpha}$ contribution when $t\rightarrow 0-$.

Deviations from the Law of the Rectilinear Diameter have indeed been observed in several careful experiments on various fluids \cite{sen:str:vic,wei:lan:for,jun:knu:hen,hen:hoh:sch:pil:fra,gol:par,pes:gol}. Among these, the seminal study of $\mbox{SF}_6$ by Weiner et al.\ \cite{wei:lan:for} more than 30 years ago reported a surprisingly sharp and abrupt singular downturn of the diameter very close to $T_c$. More specifically, their data reveal a fairly linear behavior (as predicted by the vdW equation) over a wide range of temperature up to $t \simeq -10^{-3}$; but then the diameter drops sharply to the observed critical density, $\rho_c$: see Figs.\ \ref{fig1}(a) and \ref{fig2}(a), below. This sudden drop seems to reflect the anticipated $|t|^{1-\alpha}$ singularity, provided one may assume that the Yang-Yang anomaly is negligible in this system.

Subsequent experiments on other single-component fluids, specifically, C$_2$H$_6$ and N$_2$ have revealed similar sharp singularities, while, although less definitively, they have also been observed in C$_2$H$_4$ and Ne \cite{gol:par,pes:gol}. To our knowledge, however, no theoretical or numerical calculations for model fluids have yielded such behavior. One might thus reasonably harbor some doubts about the $\mbox{SF}_6$ and similar data: see also a potential experimental concern \cite{mol:gam}. Nevertheless, Ley-Koo and Green \cite{ley:gre} analyzed these data for the diameter using the traditional restricted scaling ansatz with no pressure mixing (i.e., $j_1=j_2=0$: see below). They found that the singular $|t|^{1-\alpha}$ term with corresponding higher order corrections sufficed to describe the data for $|t| \lesssim 10^{-3}$ reasonably well with $\alpha\simeq 0.15$ (which is rather larger than the currently accepted value $\alpha \simeq 0.109$).

On the other hand, the experimentally observed coexistence curves for liquid alkali metals, such as K, Rb, Cs, etc., display quite different behavior from that of $\mbox{SF}_6$ \cite{jun:knu:hen,hen:hoh:sch:pil:fra}. These metals exhibit marked  deviations from the Law of the Rectilinear Diameter with a smooth downward curvature extending over quite a wide range of temperature even below $T\simeq 0.9\,T_c$; but they do {\em not} show a sharply dropping singularity near criticality as seen for $\mbox{SF}_6$: see Fig.\ 3, below. Clearly, in light of the varied experimental data it is of interest to see whether specific model fluids, that perhaps capture some aspects of real systems, might produce such disparate types of behavior.

Until recently, calculations for continuum (or off-lattice) model fluids, mostly computer simulations, have failed to observe singular behavior in the diameter such as seen in the experiment on $\mbox{SF}_6$ or even the extended curvature of the alkali metals. However, a newly developed scaling algorithm has enabled us to obtain precise coexistence curves for continuum models on the basis of grand canonical Monte Carlo (MC) simulations \cite{kim:fis:lui,kim:fis2}. Equipped with suitably designed finite-size scaling techniques \cite{kim:fis:lui,kim:fis}, the unbiased algorithm has generated \cite{kim} the coexistence-curve diameters up to 1 part in $10^3$ or $10^4$ of $T_c$ for a hard-core square-well (HCSW) fluid with an attractive-well of range $1.5\sigma$ (where $\sigma$ is the hard-core diameter of the particles) and for the restricted primitive model (RPM) electrolyte --- equisized hard spheres half with charge $+q$ and half with $-q$. As we report here, the simulation results for the diameter for the HCSW fluid turn out to resemble surprisingly closely the experimental data for $\mbox{SF}_6$, exhibiting a similar sharp drop very close to $T_c$! By contrast, the estimated diameter for the RPM is found to mirror the observed behavior of the liquid metals: specifically, $\rho_{\mbox{\scriptsize diam}}(T)$ curves slowly downwards towards $\rho_c$ as $T_c$ is approached with no visibly obvious singularity or rapidly growing slope. In both the RPM and the metals the role of the Coulombic interactions is felt to be paramount; so this difference in behavior, which seems also to be correlated with a large value of the Yang-Yang ratio, ${\mathcal R}_\mu$, was, perhaps, to be expected.

Our goal here is to compare quantitatively the calculated diameters of the HCSW fluid and the RPM to the experimental data for $\mbox{SF}_6$ and the liquid metals \cite{wei:lan:for,jun:knu:hen,hen:hoh:sch:pil:fra}. The comparison is instructive in that one can gauge how well these simple models capture some essential features of real fluids and, at the same time, throw light on various experimental \cite{mol:gam} and theoretical \cite{gol:ash} proposals.
\section{Theoretical background}
\label{sec2}
For completeness and perspective we sketch here some of the underlying theory. The thermodynamic potential of a one-component fluid near the critical point $(p_c,T_c,\mu_c)$ can be written in general scaling form with the leading correction as \cite{fis:ork,kim:fis:ork}
 \begin{equation}
  \tilde{p} \approx Q|\tilde{t}|^{2-\alpha}W_{\pm}(U\tilde{h}/|\tilde{t}|^\Delta,U_4|\tilde{t}|^\theta), \label{eq1}
 \end{equation}
with $\pm$ for $\tilde{t}\gtrless 0$, while, to linear order, the relevant scaling fields are
 \begin{eqnarray}
  \tilde{p} & \;=\; & \check{p} - k_0 t - l_0\check{\mu} + \cdots,  \label{eq2} \\
  \tilde{t} & \;=\; & t - l_1\check{\mu} - j_1\check{p} + \cdots,  \label{eq3} \\
  \tilde{h} & \;=\; & \check{\mu} - k_1 t - j_2\check{p} + \cdots,  \label{eq4}
 \end{eqnarray}
in which we have introduced the dimensionless critical deviations
 \begin{equation}
  \check{\mu} \equiv \frac{\mu - \mu_c}{k_{\mbox{\scriptsize B}}T_c}, \hspace{0.2in} \check{p} \equiv \frac{p - p_c}{\rho_c k_{\mbox{\scriptsize B}}T_c}, \label{eq5}
 \end{equation}
while in (\ref{eq1}) the gap exponent is $\Delta=2$$\,-\,$$\alpha$$\,-\,$$\beta=\beta$$\,+\,$$\gamma$ where, in standard notation, $\gamma$ is the exponent for the divergence of the compressibility/susceptibility, while $\theta$ is the leading correction-to-scaling exponent. Together with the exponents $\alpha$, $\beta$, $\cdots$, etc., the scaling functions $W_{\pm}(x,x_4)$ are universal. On the other hand, the amplitudes $Q$, $U$, and $U_4$ are {\em non}universal, i.e., system-dependent as are the {\em mixing coefficients} $j_1, j_2, \cdots, l_1$. In this expression we have neglected higher order corrections and nonlinear mixing terms in the scaling fields \cite{kim:fis:ork}.

The scaling ansatz (\ref{eq1})-(\ref{eq4}) accounts for the Yang-Yang anomaly: explicitly, when $j_2$ is nonvanishing, the second temperature derivative of the chemical potential on the phase boundary, $\mu_\sigma^{\prime\prime}(T)\equiv (d^{2}\mu_{\sigma}/dT^2)$, diverges at criticality like the specific heat, i.e., as $C_V \sim 1/|t|^\alpha$. The Yang-Yang ratio is then given by \cite{fis:ork,kim:fis:ork}
 \begin{equation}
  {\mathcal R}_\mu = -\lim_{T\rightarrow T_c-} T\mu_\sigma^{\prime\prime}(T)/C_V(T) = j_2/(j_2-1).  \label{eq6a}
 \end{equation}

The two sides of the coexistence curve, $\rho = \rho^{+}(T)\geq \rho_c$ and $\rho^-(T)\leq \rho_c$, follow  directly from the scaling ansatz (\ref{eq1}). The results are \cite{kim:fis:ork}
 \begin{eqnarray}
  \Delta\rho(T) & \equiv & [\rho^+(T) - \rho^-(T)]/\rho_c \nonumber \\
  & = & B|t|^\beta ( 1 + b_\theta |t|^\theta + \cdots),  \label{eq6} \\ \nonumber\\
  \Delta\check{\rho}_{\mbox{\scriptsize d}}(T) & \equiv & \frac{\rho^+(T) + \rho^-(T)}{2\rho_c} -1 \nonumber \\
  & = & A_{2\beta} |t|^{2\beta} + A_{1-\alpha}|t|^{1-\alpha} + A_1 t \nonumber \\
  & & + A_{2\beta}^{\prime}|t|^{2\beta + \theta}+ A_\theta |t|^{1-\alpha+\theta} + \cdots,  \label{eq7}
 \end{eqnarray}
where the amplitude $B$, and coefficients $b_\theta$, $A_{2\beta}$, etc., can be expressed explicitly in terms of universal derivatives of $W_\pm (x,x_4)$ and the nonuniversal values of $Q, U, U_4$ and the mixing coefficients in (\ref{eq2})-(\ref{eq4}). Note that $A_{2\beta}$ is proportional to the pressure-mixing coefficient $j_2$ \cite{kim:fis:ork}, while $A_{1-\alpha}$ is proportional to $l_1$$\,+\,$$j_1$. For the $(d$$\,=\,$$3)$ Ising universality class, which describes standard fluid criticality, one has $\alpha \simeq 0.109$ and $\beta \simeq 0.326$, as mentioned, while $\theta \simeq 0.52$ \cite{gui:zin}.

\section{Diameters of SF$_6$ and the HCSW fluid}
\label{sec3}
\begin{figure}
\centerline{\epsfig{figure=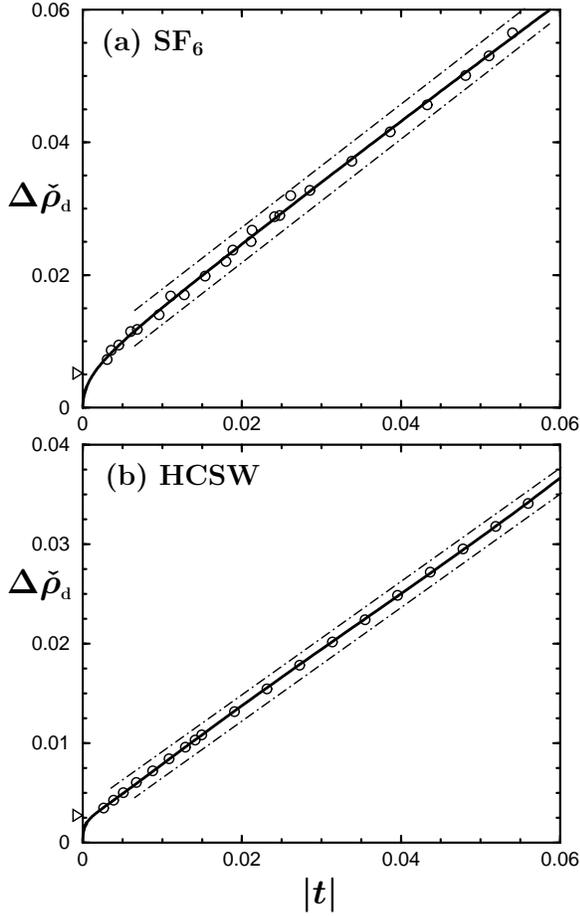,width=3.0in,angle=0}}
\caption{Plots of the reduced coexistence-curve diameters, $\Delta\check{\rho}_{\mbox{\scriptsize d}}(T)$$\,\equiv$$\,(\rho_{\mbox{\scriptsize diam}}/\rho_c)-1$ vs $t$$\, =$$\, (T-T_c)/T_c$ for (a) SF$_6$, as observed by Weiner et al.\ \cite{wei:lan:for}, and (b) the hard-core square-well (HCSW) fluid with interaction range $1.5\sigma$, as determined via Monte Carlo simulations \cite{kim}. For clarity, none of the data points closer to $T_c$ than 2-3 part in $10^3$ are shown in either plot while most of the calculated data points for $|t|<0.01$ have been omitted in (b): these data can be examined in Fig.\ \ref{fig2}. The solid curves represent fits to all the data as detailed under (ii) in Table \ref{tab1}. Conversely, the dot-dashed lines that narrowly encompass the data points serve to demonstrate that the Law of the Rectilinear Diameter is well obeyed for $t\geq 0.005$. Note, however, that this linear behavior extrapolates to offsets, $\Delta\rho_c$, {\em above} $\rho_c\,$, as marked by the open arrowheads on the ordinate scales. \label{fig1}}
\end{figure}

The coexistence-curve diameter of $\mbox{SF}_6$ measured by Weiner et al.\ \cite{wei:lan:for} is presented in Fig.\ \ref{fig1} together, 
in order to make a comparison, with the simulation results for the HCSW fluid \cite{ork:fis:pan,kim}. Both observation and calculation span almost three orders of magnitude in temperature reaching up to within a few parts in $10^5$ of $T_c$. It is striking, first, to see that the HCSW fluid, the most basic continuum model that undergoes gas-liquid phase separation, exhibits a behavior that so closely matches that for $\mbox{SF}_6$. Indeed, the diameters of both systems show essentially linear behavior for $|t|\gtrsim 0.005$, as may be gauged by the pair of parallel straight, dot-dashed lines that embrace the data in both Figs.\ \ref{fig1} (a) and (b). Note, however, that extrapolation of these rectilinear regimes to criticality yields density offsets, $\Delta\rho_c$, above the true value of $\rho_c$ amounting to about $0.53\%$ for $\mbox{SF}_6$ and $0.23\%$ for the HCSW fluid. Then, the diameters for both systems display unexpectedly sharp downwards departures from linearity that set in around $t = -(3$-$5)$$\,\times\,$$10^{-3}$ and suggest a divergent derivative {\em at} $T_c$. As far as we are aware this is the first time that such seemingly singular behavior in the diameter of a model fluid has been identified via computer simulations. 

\begin{figure}
\centerline{\epsfig{figure=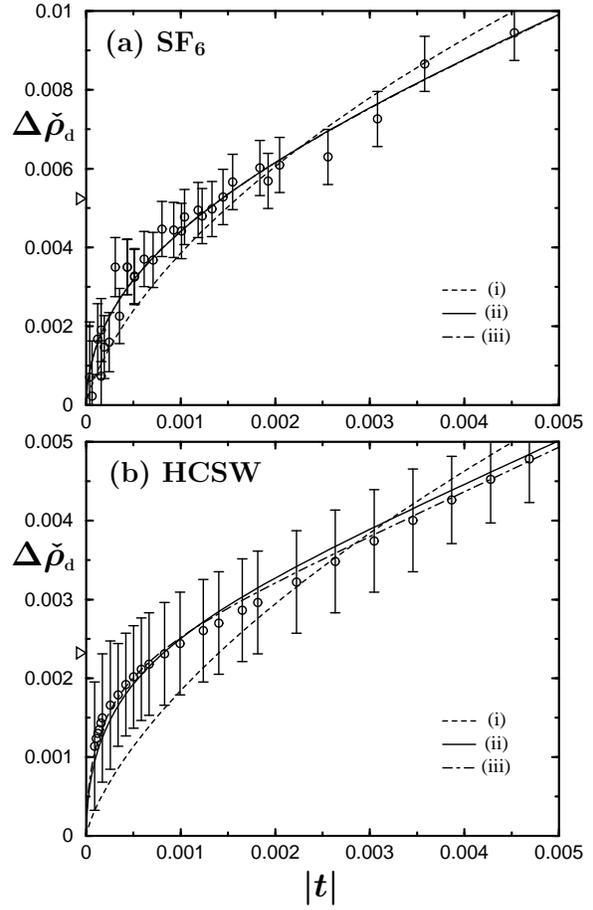,width=3.0in,angle=0}}
\caption{The reduced coexistence-curve diameters, $\Delta\check{\rho}_{\mbox{\scriptsize d}}(T)$, close to $T_c$ for (a) SF$_6$ \cite{wei:lan:for} and (b) for the HCSW fluid \cite{kim}. The SF$_6$ data presented here [and in Fig.\ \ref{fig1}(a)] are reproduced from Figs.\ 1 and 2 of Weiner et al.\ \cite{wei:lan:for} (with neglect of the two data points closest to $T_c$ that were subject to systematic displacements entailed in linear extrapolations for the gravitational corrections implemented). The various curves represent fits to the data based on the expression (\ref{eq7}) with the coefficients given in Table \ref{tab1}. Note that the dot-dashed plot (iii) in (a) is almost identical to the solid curve, the differences being visible only for $t\gtrsim 0.003$. \label{fig2}}
\end{figure}

In order to reveal the putative singularities in more detail, Fig.\ \ref{fig2} 
presents magnified plots of the diameters over the much smaller temperature interval $|t|\leq 0.005$. Although the diameter of the HCSW fluid exhibits a somewhat sharper drop, the quantitative similarity in this {\em non}universal aspect of the critical behavior is surprising granted the crude nature of the HCSW fluid regarded as a model of a real system undergoing gas-liquid phase separation. The close correspondence between the observations and the theory suggests that the issue raised by Moldover and Gammon \cite{mol:gam} (and cited in \cite{gol:ash}; but see also \cite{pes:gol}) regarding the possible influence of wetting layers on the observations of Weiner et al.\ should {\em not} be regarded as a serious concern.

In order to understand the contributions of the singular terms to the diameter, we have examined fits of the data to (\ref{eq7}) subject to various conditions. First, we may neglect the $|t|^{2\beta + \theta}$ term, since its amplitude $A_{2\beta}^{\prime}$ is proportional to $j_2$ and will probably be of small if any significance in light of the estimated and likely value of the Yang-Yang ratio, ${\mathcal R}_\mu$ \cite{fis:ork,kim}. Next, on ignoring, initially, the leading term varying as $|t|^{2\beta}$, that originates from the pressure mixing (and so is also proportional to ${\mathcal R}_\mu$), we find least-squares fits that seem to be significantly {\em less than satisfactory}: see the dashed lines labeled (i) in Fig.\ \ref{fig2}. The associated values of the coefficients for $\mbox{SF}_6$ are about twice those for the HCSW fluid: see lines (i) in Table \ref{tab1}. 
\begin{table}
\caption{Parameters for fitting the diameters of SF$_6$ \cite{wei:lan:for} and the HCSW fluid \cite{kim} to Eqns.\ (\ref{eq7}) and (\ref{eq8}). \label{tab1}}
\begin{tabular*}{\linewidth}{@{\extracolsep{\fill}}c|c|ccccc}\hline\hline
  & & $A_{2\beta}$ & $A_{1-\alpha}$ & $A_1$ & $A_\theta$ & $a_\theta$ \\ \hline
 (i) & SF$_6$ & ---  & 6.365 & $-10.13$ & 8.080 & --- \\
  & HCSW$~~$ & --- & 2.897 & $-4.532$ & 3.884 & --- \\ \hline
 (ii) & SF$_6$ & 1.124 & $-9.042$  & 11.37 & $-3.354$ & --- \\
  & HCSW & 0.965 & $-9.818$ & 12.97 & $-5.092$ & --- \\ \hline
 (iii)$~$ & SF$_6$ & 1.086 & $-7.990$  & 9.770 & --- & 3.318 \\
  & HCSW & 1.143 & $-11.15$ & 14.38 & --- & 12.02\\ \hline\hline
\end{tabular*}
\end{table}
Of course, this might well have been anticipated from Fig.\ \ref{fig2} in which the vertical scale for $\mbox{SF}_6$ is twice that for the HCSW fluid. Nevertheless, the fact that the ratio of corresponding coefficients lies within the rather narrow range $2.16\pm 0.08$ provides an interesting measure of the similarity of the two data sets.

On the other hand, if one also allows the leading, pressure mixing term in (\ref{eq7}), the fits are markedly improved: see the solid curves in Figs.\ \ref{fig1} and \ref{fig2}. Furthermore, the values of the three leading fitted coefficients are now surprisingly close for the two systems; indeed, only the last amplitude, $A_\theta$ (which pertains to the leading corrections to scaling) differs by more than $17\,\%$: see (ii) in Table \ref{tab1}. 

To gain some insight as to what might be reasonable confidence limits to attach to these results, we have fitted the data to the generalized Pad\'{e} approximant form
 \begin{equation}
  \Delta\check{\rho}_{\mbox{\scriptsize d}} \simeq \frac{A_{2\beta}|t|^{2\beta} + A_{1-\alpha}|t|^{1-\alpha} + A_1 t}{1 + a_\theta |t|^{\theta}}, \label{eq8}
 \end{equation}
where the $|t|^\theta$ term in the denominator accounts for infinitely many higher order contributions (including the neglected $|t|^{2\beta + \theta}$ term). The results of this fit are presented in lines (iii) in Table \ref{tab1} and as dot-dashed curves in Fig.\ \ref{fig2}. As seen in the figure, the differences between the fits (ii) and (iii) are negligible. Furthermore, the values of the three leading coefficients have changed by no more than $20\%$.

Now, according to the scaling ansatz (\ref{eq1})-(\ref{eq4}), the leading amplitude in (\ref{eq7}) can be expressed as \cite{kim:fis:ork}
 \begin{equation}
  A_{2\beta} = {\mathcal R}_\mu B^2, \label{eq9}
 \end{equation}
where ${\mathcal R}_\mu$ is the Yang-Yang ratio defined as in (\ref{eq6a}). This ratio and the amplitude $B$ have been estimated rather precisely for the HCSW fluid \cite{kim:fis:lui,kim}. Using the values ${\mathcal R}_\mu \simeq -0.04$ and $B\simeq 2.0$ then yields $A_{2\beta}\simeq -0.16$. This value, however, is not only an order of magnitude smaller than the fitted values in Table \ref{tab1}, but also has the opposite sign! This indicates unequivocally that the fitted coefficients are unlikely to represent the true amplitudes for the diameter. Possible reasons are, first, that since the contributing terms in (\ref{eq7}) have exponents that are closely spaced numerically, they interfere strongly in the fitting procedure, resulting in {\em effective} rather than realistic amplitudes. Second, one knows that there are higher order terms which have been totally ignored in the fitting but that might play a crucial role. One such is the leading odd correction-to-scaling contribution varying as $|t|^{\beta+\theta_5}$ \cite{kim:fis:ork}; others will arise from the nonlinear contributions to the scaling fields (\ref{eq2})-(\ref{eq4}) \cite{kim:fis:ork}. Finally, if one fixes $A_{2\beta}$ at the expected value $-0.16$ and considers the approximant (\ref{eq8}) with an extra $A_\theta^\prime |t|^{1-\alpha+\theta}$ term in the numerator, one finds that the diameter of the HCSW fluid can be described just as well with the coefficients $A_{1-\alpha}\simeq 206$, $A_1 \simeq -377$, $A_\theta^\prime \simeq 1144$ and $a_\theta \simeq 1872$ as with the sets (ii) and (iii) in Table \ref{tab1}. From this lesson, we learn that beyond demonstrating the close similarity of the two systems, one must not attach much significance to the actual coefficient values for $\mbox{SF}_6$ and the HCSW fluid in Table \ref{tab1}. In order to obtain reliable estimates for even the two leading singular amplitudes, $A_{2\beta}$ and $A_{1-\alpha}\,$, one would require measurements of the diameter with markedly higher precision and still closer to $T_c$.

Finally, we mention that Goldstein and coworkers \cite{gol:par,pes:gol} examined the trends with molecular properties of both the slopes of the coexistence-curve diameters and the strengths of the singularity for `normal fluids,' \cite{sin:pit} such as SF$_6$, C$_2$H$_6$, N$_2$, etc. They argued that a crucial role is played by {\em triplet} or {\em three-body} intermolecular interactions that, in turn, may be quantified by the molecular polarizabilities \cite{gol:par,pes:gol}. (Observed diameter slopes had previously been correlated with quantum-mechanical effects as embodied in the De Boer parameter $\Lambda^\ast {\,=\,} h/\sqrt{m\sigma^2 \epsilon}$, where $m$, $\sigma$ and $\epsilon$ represent the molecular mass, collision diameter and attractive well depth, respectively \cite{fis}.) In contradiction, however, later measurements for xenon \cite{nar:bal}, for which the polarizability product, $\alpha_p\rho_c$ \cite{gol:par,pes:gol} is unusually high, reveal a diameter of large slope but undetectably small deviations from rectilinearity close to $T_c$. Furthermore, Singh and Pitzer \cite{sin:pit} studied a larger array of normal fluids and concluded generally that the slope of the diameter does {\em not} correlate with three-body forces, but, rather, is primarily determined by the shape of the two-body potential. That, in turn, relates closely to the {\em acentric factor} defined via the slope of the vapor pressure curve at $T{\,=\,} 0.7\,T_c$ \cite{sin:pit}. In as far as there are no many-body interactions in HCSW models, our results for the singularity are in accord with this conclusion.

\section{Diameters of liquid metals and the RPM}
\label{sec4}
\begin{figure}
\centerline{\epsfig{figure=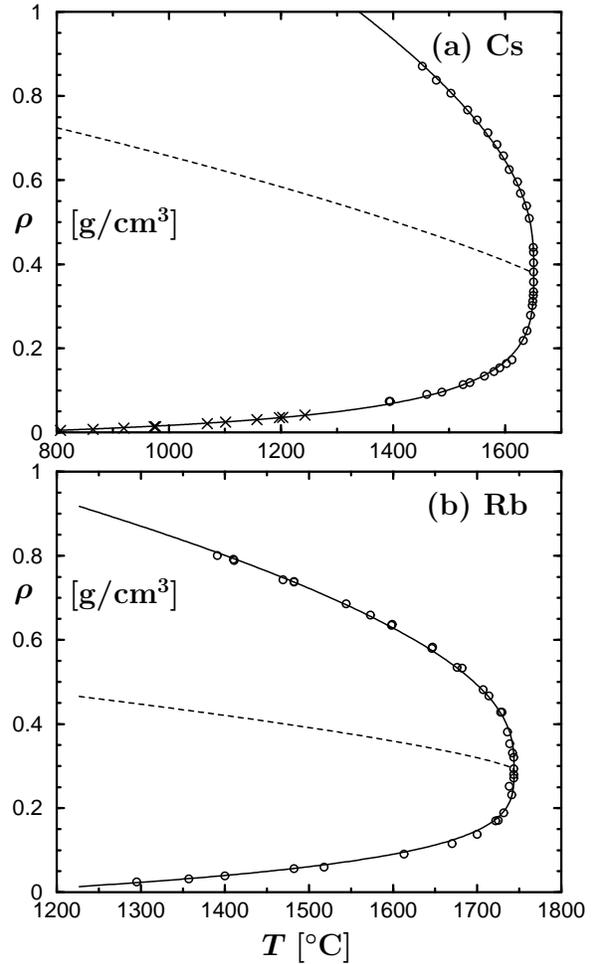,width=3.0in,angle=0}}
\caption{The coexisting densities of Cs and Rb as reported by Hensel and coworkers \cite{jun:knu:hen} who invoked further data for Cs (crosses) from Stone et al.\ \cite{sto}. The solid curves portray fits to (\ref{eq6}) and (\ref{eq7}) while the dashed curves represent the corresponding diameters: see Table \ref{tab2} for the coefficients employed. Note that the tabulated coexistence curves of these liquid alkali metals \cite{jun:knu:hen} were obtained by extending isochores down to the previously observed vapor pressure curves. Since, as a result, the liquid and vapor densities are available only at differing temperatures, the diameters cannot be determined directly and fitted only to (\ref{eq7}). \label{fig3}}
\end{figure}

The coexistence curves of the liquid alkali metals \cite{jun:knu:hen,hen:hoh:sch:pil:fra} are found to exhibit a large degree of asymmetry and so are quite distinct in their overall shapes from those of SF$_6$ and other normal fluids and from the HCSW model. See Fig.\ \ref{fig3} 
for Cs and Rb and Hensel et al.\ \cite{hen:hoh:sch:pil:fra} for Na and K. This behavior can be compared to that of the RPM \cite{kim:fis:lui,kim,lui:fis:pan,kim:fis3} which is even more strongly asymmetric. The relative similarity of the coexistence curves of the liquid metals and the RPM would appear to be instructive since the long-range Coulombic interactions play a dominant role in both types of system. 

In order to obtain the diameters of the liquid metals, we have performed a global least-squares fit to the coexisting densities, $\rho^{\pm}(T)$, on the basis of the expansions (\ref{eq6}) and (\ref{eq7}). The values of $T_c$ and $\rho_c$ used in our fits for Cs and Rb are those stated by Hensel and coworkers \cite{jun:knu:hen} (But see further below). The resulting amplitudes and coefficients are presented in Table \ref{tab2} 
\begin{table}
\caption{Parameters for fitting the liquid and vapor densities of the alkali metals \cite{jun:knu:hen,sto} and the RPM \cite{kim:fis:lui,kim} to Eqns.\ (\ref{eq6}) and (\ref{eq7}). \label{tab2}}
\begin{tabular*}{\linewidth}{@{\extracolsep{\fill}}c|cccccc}\hline\hline
   & $A_{2\beta}$ & $A_{1-\alpha}$ & $A_1$ & $A_\theta$ & $B$ & $b_\theta$ \\ \hline
  Rb & $-5.041$ & 54.65  & 62.73 & 17.21 & 1.908 & 0.556 \\ \hline
  Cs & $-0.232$ & $2.206$ & $-0.611$ & $-0.911$ & 1.927 & 0.436 \\ \hline
 RPM$~$ & $-3.120$ & 45.85  & 47.09 & 9.564 & 6.932 & 0.312 \\ \hline\hline
\end{tabular*}
\end{table}
together with the corresponding RPM values. The coefficients $A_{2\beta}$, $\cdots$, $A_\theta$ for the diameter of Rb prove quite comparable to those of the RPM; however, the leading amplitude, $B$, for the density discontinuity, $\Delta\rho(T)$, of the RPM is larger by a factor close to 3.6 which simply reflects the markedly greater asymmetry of the model relative to the liquid metals and, concomitantly, also entails a more steeply sloping diameter \cite{kim,lui:fis:pan}.

\begin{figure}
\centerline{\epsfig{figure=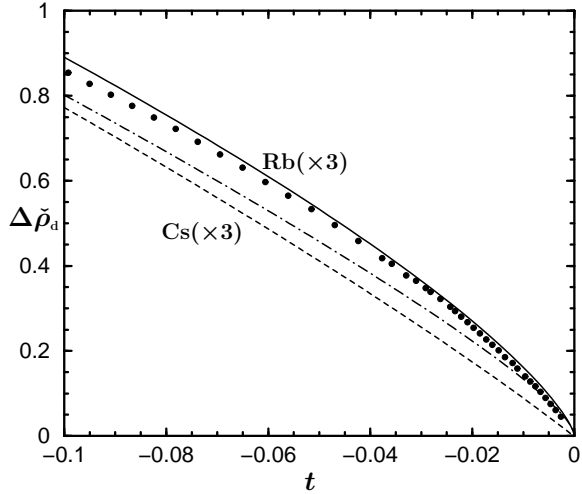,width=3.0in,angle=0}}
\caption{The reduced coexistence-curve diameters, $\Delta\check{\rho}_{\mbox{\scriptsize d}}(T)$, of Cs (dashed), Rb (solid curve), and the RPM (solid circles). The plots for Cs and Rb, derived from \cite{jun:knu:hen} by fitting to (\ref{eq6}) and (\ref{eq7}), have been amplified by a factor 3 to facilitate comparison with the calculations for the RPM for which the uncertainties are judged to be no larger than twice the symbol size \cite{kim}. The dot-dashed curve for Cs is found by changing slightly (within acceptable uncertainty limits) the stated values for $T_c$ and $\rho_c$ \cite{jun:knu:hen}: see text. \label{fig4}}
\end{figure}

The diameters of the liquid metals determined by these procedures and of the RPM (as found in \cite{kim}) are presented in Fig.\ \ref{fig4}. 
It is striking that, apart from the anticipated multiplication factor close to 3, the diameter of the RPM is almost identical to that of Rb. Both diameters exhibit smoothly varying curvature down to $t\simeq -0.2$ or lower and differ significantly from those of SF$_6$ and the HCSW fluid. The diameter of Cs exhibits a less pronounced curvature, but still shows distinctive departures from rectilinear behavior as, indeed, originally stressed by Hensel and coworkers. It should be noted, however, that if one allows small deviations from the quoted critical parameters \cite{jun:knu:hen}, namely, say $\sim 0.02\%$ of $T_c$ (or only $\lesssim 0.4\,$K) and of $1\%$ in $\rho_c$ (which seems allowable in light of the tabulated data), the revised fits reproduce curvatures as large as those for Rb: see the dot-dashed plot in Fig.\ \ref{fig4}.

Just as for the HCSW fluid, the leading amplitude, $A_{2\beta}$, of the diameter for the RPM can be estimated via (\ref{eq9}) by using the value ${\mathcal R}_\mu \simeq 0.26$ \cite{kim:fis:lui,kim} and the amplitude $B$ given in Table \ref{tab2}: this yields $A_{2\beta}\simeq 12.5$ which is four times the magnitude of the fitted coefficient in Table \ref{tab2}, and, more disturbingly, has the opposite sign. Once again, this demonstrates that fitted amplitudes, such as recorded in Table \ref{tab2}, may not be regarded as realistic. At best, owing to the interplay of closely spaced exponent values and contributions from higher order corrections, the data in Table \ref{tab2} must be regarded as effective values that, {\em collectively}, describe the observations reliably. This is further supported by the fact that fixing $A_{2\beta}$ to the theoretical value $12.5$ and using the approximant (\ref{eq8}) yields a satisfactory fit with the coefficients $A_{1-\alpha} \simeq -199$, $A_1 \simeq -327$ and $a_\theta \simeq 35$.

Two further remarks are called for. First, it must be remembered that the sudden downward turns of the diameters of SF$_6$ and the HCSW fluid set in around $t=-(2$-$3)$$\,\times\,$$10^{-3}$ and are clearly developed only for $|t|\lesssim (4$-$5)$$\,\times\,$$10^{-4}$: see Figs.\ \ref{fig1} and \ref{fig2}. Owing to experimental difficulties, however, reliable data for the alkali metals have been obtained only for $|t|\gtrsim 10^{-3}$ \cite{jun:knu:hen}. It is possible, therefore, that further observations approaching $T_c$ by another decade or more would also uncover a sharp drop in the liquid metal diameters very close to $T_c$ ({\em in addition} to the extended curved region further from criticality).

Second, the experiments of Hensel and coworkers were associated with an interesting theoretical suggestion of Goldstein and Ashcroft \cite{gol:ash}. They argued that the characteristic curvature of the coexistence-diameters of the liquid metals (in contrast to the largely rectilinear behavior observed for typical insulating or normal fluids) ``arises from correspondingly strong thermodynamic-state dependence of the screened ion-ion interactions" in such systems. In addition, they suggested an association with the crossing of the metal-insulator transition. As regards this second proposal, we believe the weight of evidence from various studies now suggests that the metal-insulator {\em crossover} (which is {\em not} a sharp transition at nonzero temperatures) has essentially no correlations with thermodynamic properties beyond residing in the wide range of fluid densities accessible near and above any gas-liquid critical point: see, e.g., studies of the phase behavior of liquid mercury through criticality \cite{koz}.

However, the close similarity of the coexistence diameters of the liquid metals and the restricted primitive model (RPM) electrolyte casts a fresh light on the first suggestion. In as far as the RPM is a perfectly standard  statistical mechanical model with well-defined {\em fixed}, i.e., {\em state}-{\em independent} interaction potentials, albeit of long range, the focus of Goldstein and Ashcroft on state-dependent potentials {\em per se} might seem inappropriate or, at least, unnecessary. Certainly, neither quantum mechanics, strongly delocalized electronic states, or rapid changes from localized tight-binding to itinerant band states, nor the associated electron-ion mass and interaction disparities can be directly relevant!

On the other hand, if one wishes to view a classical ionic systems as an uncharged fluid with effective ``screened interactions" one is naturally led to consider pair potentials of Debye-H\"{u}ckel form. The interaction range is then set by the charge screening length, $\xi_Z(T,\rho)$, which is certainly state-dependent, approaching the Debye value, $\xi_D \propto (T/\rho)^{1/2}$ at low density and high temperatures \cite{bek:fis}. However, Goldstein and Ashcroft \cite{gol:ash} characterize such screening in ``typical electrolyte solutions" as only ``weakly state dependent" in contrast to the ``vast qualitative change in the form of the interaction which occurs at the metal-insulator transition." Although theoretical arguments \cite{aqu:fis} indicate that, even in the fully charge-symmetric RPM, weak energy-like singularities appear in various charge correlation lengths near criticality, it is hard, in our opinion, to view these changes, even if relatively rapid, as {\em driving} the curvature and seeming singular behavior of the coexistence-diameter in the RPM {\em or}, mutatis mutandis, in the liquid metals.

Nevertheless, the Coulomb dominated behavior of the RPM and of the liquid metals does appear to be directly associated with the coexistence-curve asymmetries and, it even seems plausible, with relatively large magnitudes of the Yang-Yang ratio ${\mathcal R}_\mu$. Further experiments, simulations, and calculations to investigate these surmises would surely be of interest.
\section{Summary}
\label{sec5}
In summary, the experimentally observed singular behavior or, more precisely, the deviations from rectilinear variation of the coexistence-curve diameters of both SF$_6$ and liquid alkali metals have been compared with rather precise numerical calculations for a hard-core square-well (HCSW) fluid and for the restricted primitive model (RPM) electrolyte \cite{kim}. The close similarity between the accurately measured diameter of SF$_6$ \cite{wei:lan:for} and that of the HCSW fluid is remarkable; the sharp singularity setting in very close to $T_c$ is striking and resembles the entropy-like singularity proposed long ago even though fits to the data are unable to confirm the specific analytic expectations unequivocally. By contrast, the markedly curved diameters discovered for the liquid alkali metals \cite{jun:knu:hen} exhibit very different behavior from those of SF$_6$ and the HCSW fluid; it transpires, instead, that they closely reflect the calculated behavior for the RPM. While it seems likely that the differences from the non-conducting systems are intimately associated with the long-range Coulomb interactions, proposals \cite{gol:ash} emphasizing the delocalized, quantum-mechanical electronic states normally used to describe liquid metals, do not seem pertinent.

Finally, let us mention that the rapidly dropping singular behavior seen in the coexistence-curve diameters for the HCSW fluid and for SF$_6$ and various single-component fluids has not so far been observed in simulations of other models. A similar sharp near-critical behavior has been reported in the diameter of a binary-liquid system \cite{gop:ram:cha} and, indeed, the same spectrum of singularities is to be expected in general. However, because of the extra thermodynamic degree of freedom in a binary system, a further set of mixing coefficients for the second chemical potential must be invoked. As a result both the theoretical scaling analysis and the quantitative elucidation of experimental observations face extra challenges. Nevertheless, further experimental studies of such systems and, indeed, corresponding calculations could be instructive.

{\bf Acknowledgments}

The authors are grateful to Jan Sengers and Mikhail Anisimov for their interest and indebted to Benjamin Widom, Raymond Goldstein, and, especially, Neil Ashcroft for comments on a draft manuscript and related topics. The support of the National Science Foundation (through Grant No.\ CHE 03-01101) is gratefully acknowledged.

\end{document}